\documentclass[twocolumn]{aastex61}
\usepackage{natbib,twoopt,amsmath}

\received{January 16, 2017}
\revised{February 9, 2017}
\accepted{February 10, 2017}

\submitjournal{ApJ}

\shorttitle{3D current structures}
\shortauthors{Nickeler et al.}

\begin{document}

\title{Electric current filamentation induced by 3D plasma flows in the solar corona}

\correspondingauthor{Dieter H. Nickeler}
\email{dieter.nickeler@asu.cas.cz}

\author{Dieter H. Nickeler}
\affil{Astronomick\'y \'ustav, Akademie v\v{e}d \v{C}esk\'e republiky, v.v.i., Fri\v{c}ova 298, 251\,65 Ond\v{r}ejov, Czech Republic}

\author{Thomas Wiegelmann}
\affiliation{Max-Planck Institut f\"{u}r Sonnensystemforschung, Justus-von-Liebig-Weg 3, 37077 G\"{o}ttingen, Germany}

\author{Marian Karlick\'y}
\affiliation{Astronomick\'y \'ustav, Akademie v\v{e}d \v{C}esk\'e republiky, v.v.i., Fri\v{c}ova 298, 251\,65 Ond\v{r}ejov, Czech Republic}

\author{Michaela Kraus}
\affiliation{Astronomick\'y \'ustav, Akademie v\v{e}d \v{C}esk\'e republiky, v.v.i., Fri\v{c}ova 298, 251\,65 Ond\v{r}ejov, Czech Republic}
\affiliation{Tartu Observatory, 61602 T{\~o}ravere, Tartumaa, Estonia}

\begin{abstract}
Many magnetic structures in the solar atmosphere evolve rather slowly so that they can be 
assumed as (quasi-)\-static or (quasi-)stationary and represented via magneto-hydrostatic (MHS)
or stationary magneto-hydrodynamic (MHD) equilibria, respectively. While exact 3D solutions 
would be desired, they are extremely difficult to find in stationary MHD. 
We construct solutions with magnetic and flow vector fields that have three components 
depending on all three coordinates.
We show that the non-canonical transformation method produces quasi-3D solutions of stationary MHD
by mapping 2D or 2.5D MHS equilibria to corresponding stationary-MHD states, i.e., states that display 
the same field line structure as the original MHS equilibria.
These stationary-MHD states exist on magnetic flux surfaces of the original 2D MHS-states.
Although the flux surfaces and therefore also the equilibria have a 2D character, these
stationary MHD-states depend on all three coordinates and display highly complex currents.
The existence of geometrically complex 3D currents within symmetric field-line structures
provide the base for efficient dissipation of the magnetic energy in the solar corona by Ohmic heating.
We also discuss the possibility of maintaining an important subset of
non-linear MHS states, namely force-free fields, by stationary flows. We find that force-free fields 
with non-linear flows only 
arise under severe restrictions of the field-line geometry and of the magnetic flux 
density distribution.

\end{abstract}


\section{Introduction}

Many structures in the atmosphere of the sun and in solar-type stars evolve on relatively large
time-scales so that they can be described within the frame of quasi-stationary or quasi-static
magneto-hydrodynamics (MHD). Prominent examples are solar arcade structures, loops as well as
prominences. For their representation typically magneto-hydrostatic (MHS) equilibria 
\citep[e.g.,][]{1982ApJ...263..952L, 2015AstL...41..211S} or stationary-state models, i.e., 
stationary MHD equilibria, are calculated \citep[e.g.,][]{1999GApFD..91..269P, 2005A&A...429.1081P}.

Generally, it would be desirable to have a full 3D representation of the MHD equilibrium states.
However, as was already mentioned by \citet{1972ApJ...174..499P}, it is normally not possible
to construct 3D states in the functional vicinity of 2D states. This means that 2D equilibria on which 
perturbations are imposed typically do not relax into smooth 3D states \citep{1982ApJ...259..832T}.
Instead, the resulting equilibria must contain tangential discontinuities, i.e., singular currrent sheets. 
This is known as Parker's conjecture \citep{1983ApJ...264..642P, 1983ApJ...264..635P, 1988ApJ...330..474P}
which states that no regular equilibria exist without a symmetry. In this context, symmetry does not 
necessarily imply that the system has an ignorable coordinate \citep{1985SoPh..100..309L}, where 
ignorable coordinate means that in a specific coordinate system the physical values do not depend on 
this coordinate. We note that under specific circumstances a few regular classes of 3D MHS states have been found
\citep[see, e.g.,][]{1982ApJ...263..952L, 1995A&A...301..628N}, and a set of exact analytical 3D stationary
MHD flows exists as well \citep[see, e.g.,][]{2001PhLA..291..256B, 2002PhRvE..66e6410B}, however, the computation
of these solutions requires that a complete stationary flow must already be known. 

According to Parker, the appearance of singular current sheets could provide a suitable mechanism 
for acceleration and heating of the coronal plasma via Ohmic heating, i.e., Joule dissipation, caused by 
magnetic reconnection within these current sheets. To guarantee that heating is provided on a regular base
(i.e. also during times with no huge eruptions), successive heating should take place. This can only
be achieved considering quasi-continuous small scale eruptions, the so-called nanoflares \citep{1988ApJ...330..474P}.    
However, it is still highly debated whether large-scale eruptions or small-scale nanoflares are the major         
mechanism for the heating of the solar coronal plasma \citep[see, e.g.,][]{2012RSPTA.370.3217P, 
2016ApJ...831....9S}.

Shearing motions of the magnetic field lines, e.g. at the footpoints of arcade structures, can be 
used to produce nanoflares \citep[see, e.g.,][]{2011A&A...530A.112B, 2013A&A...555A.123B, 2015ApJ...811..106H} or 
large-scale eruptions \citep[e.g.,][]{2003JGRA..108.1162M, 2013SoPh..284..447K, 2013ApJ...773..128T,
2014ApJ...787...46L}. Such a procedure does not necessarily converge into an equilibrium state anymore. 
Therefore, in numerical simulations
these sheared field lines might be forced to relax into an equilibrium state by introducing numerical
resistivities and viscosities \citep[e.g.,][]{2011A&A...536A..67W, 2016A&A...587A.125P}.

Another approach for small-scale eruptions and heating was made by \citet{2015MNRAS.454.1503P}, who applied 
a so-called volumetric Parker model. This model is not based on the shearing motions of the footpoints.
Instead, a large-scale 
motion of the magnetic field lines is applied throughout the volume of the fluid. This large-scale 
motion is driven by an initial stationary flow, generated by a time-dependent stream function whose Fourier 
components are kept fixed at each time step. These stationary flows generate additional turbulent flows, 
which are allowed to evolve in time. 

Alternatively, a model including selfconsistent plasma flows was developed by
\citet{2013A&A...556A..61N, 2014A&A...569A..44N}. 
This model produces highly fragmented, strongly peaked 
currents and vortices spreading from large to small scales, while the system remains in a well-defined 
equilibrium.

In most of the aforementioned approaches, the initial condition is either a static or some arbitrary
field that is non compatible with the resulting flow field. The numerically calculated corresponding 
changes of the fields are, therefore, based either on linear or non-linear perturbation theory or on stochastics. 
What is often neglected is that observations imply stationary flows in active regions and coronal holes rather 
than pure force-free or static fields \citep{2001ApJ...553L..81W, 2002ApJ...567L..89W, 2004A&A...428..629M, 
2005A&A...432L...1W}. Also, during pre-flare stages upflows in the photosphere and 
flows along loops were observed \citep[e.g.,][]{1971PASJ...23..443Y, 1976SoPh...47..233H, 2010SoPh..267..361W}. 
Hence, an initial condition including stationary flows, as was presented by \citet{2010AnGeo..28.1523N, 
2012AnGeo..30..545N} seems more appropriate.  

Non-linear MHD flow models for loops, sun spots, and magnetic arcade structures exist 
\citep[see, e.g.,][]{1993A&A...275..613T, 1999GApFD..91..269P, 2002A&A...382.1081P, 2005A&A...429.1081P}, 
however, they were not developed explicitly for the purpose of explaining coronal heating. 
Nevertheless, non-linear MHD theory provides the proper tool for particle acceleration via generation of 
electric fields in a slightly non-ideal/resistive environment, and, therefore, for local heating processes 
\citep{2014A&A...569A..44N}.

In this paper, we wish to reinforce Parker's conjecture of heating via multiple current sheets and multiple
reconnection sites. In connection with the equilibrium problem introduced by \citet{1972ApJ...174..499P} we need 
therefore a proper method that allows slight deviations from symmetric 2D to (almost) 3D structures.
The known magnetic flux densities and the corresponding 
derived currents obtained from observations are far below the threshold for sufficient dissipation of 
magnetic energy in general, i.e. Joule heating by extremely strong currents in the case of e.g. Spitzer 
resistivity, and/or the threshold for anomalous resistivity triggering
magnetic reconnection. This implies that the current density on these scales is too low to produce
current-driven micro-instabilities. However, the observed large-scale fields might display steep gradients
on smaller scales. Complex flow patterns and steep gradients in active regions indicate the existence of shear 
flows, as was reported by \citet{2004A&A...428..629M}. The changing of the magnetic field structure often 
seems to coincide with sharp changes in the flows. Hence, this trend might be expected to continue when going 
to even smaller, yet unresolved scales. 
 
For a better comprehension of Ohmic heating and acceleration of plasma and particles, we need more detailed
information about current sheet structures in the solar atmosphere. While both observations and numerical simulations
currently cannot resolve small-scale structures, an analytical approach is a useful physical approximation that 
provides detailed information down to the theoretical dissipation scales, which are for solar corona conditions below 10 m. 
Based on the non-canonical transformation 
method developed by \citet{1992PhFlB...4.1689G} and utilized by e.g. \citet{2015PhPl...22b2520C}, we 
will show that there is a connection between the breaking of the symmetry and the 
down-cascading of the current sheet scales.
The breaking of the symmetry is done by 
field-aligned flows which have a strong gradient perpendicular to the 
field lines. These flows cause strong strong gradients of the magnetic field strength normal to the 
field lines, implying small-scale current sheets.

\section{Problem description and basic assumptions}

The magnetic field structures in the solar atmosphere, especially in the corona,
resemble magnetic arches and also closed field line structures emulating flux ropes, surrounded 
by bundles of open field lines. These magnetic structures form the stage on which chromospheric and coronal 
heating takes place. For a reasonable representation of these structures, it is necessary to calculate 
the non-linear fields forming the magnetic scaffold in the frame of stationary MHD.

\subsection{Stationary MHD equations}

We focus on incompressible field-aligned sub-Alfv\'enic flows, because they 
are exclusively related to MHS states.  This has been proved by \citet{1992PhFlB...4.1689G}, 
\citet{2006A&A...454..797N}, and \citet{2010AnGeo..28.1523N}, who found that only incompressible field-aligned 
MHD flows can be unambiguously reduced to MHS-type equations. MHS equilibria are therefore 
an infinitesimal small subset of field-aligned incompressible flows. 

Another advantage of field-aligned flows is that they guarantee that, according to ideal Ohm's law, 
the electric field in ideal MHD vanishes
\begin{equation}
\textbf{\textit{E}} +  \textbf{\textit{v}} {\boldsymbol\times} \textbf{\textit{B}} =  \textbf{\textit{0}} \quad
\Rightarrow \quad  \textbf{\textit{E}} =  \textbf{\textit{0}}\, ,
\end{equation}
and, therefore, fulfills automatically the condition that the electric field is stationary.

To simplify the representation of the equations
we introduce normalized parameters. These require the definition of 
normalization constants $\hat{B}, \hat{\rho}, \hat{l},
\hat{v}_{A}$, and $\hat{p}$. 
Let $\textbf{\textit{v}}$ be the plasma velocity normalized by the normalized Alfv\'en velocity
$\hat{v}_{A}=\hat{B}/\sqrt{\mu_{0}\hat{\rho}}$,
$\rho$ the mass density normalized by $\hat{\rho}$,
$\textbf{\textit{j}}={\boldsymbol\nabla\boldsymbol\times} \textbf{\textit{B}}$ the current density vector 
normalized by
$\hat{B}/(\mu_{0}\hat{l})$ with $\hat{l}$ as the characteristic length scale, and
$p$ the scalar plasma pressure \textbf{normalized by}
$\hat{p}=\hat{B}^2/\mu_{0}$.
Hence, the set of equations of stationary, field-aligned
incompressible MHD, consisting of the mass continuity equation, the Euler equation, the
definition for field-aligned flow and Alfv\'{e}n Mach number $M_{A}$, the incompressibility condition,
and the solenoidal condition for the magnetic field, can be written in the following form
\begin{eqnarray}
{\boldsymbol\nabla\boldsymbol\cdot}(\rho \textbf{\textit{v}}) & = & 0\, ,\label{konti}\\
\rho\left(\textbf{\textit{v}}{\boldsymbol\cdot}{\boldsymbol\nabla}\right)\textbf{\textit{v}} & = & \textbf{\textit{j}}{\boldsymbol\times}\textbf{\textit{B}} - {\boldsymbol \nabla} p\, ,
\label{euler0} \\
\textbf{\textit{v}} & = & \frac{M_{A}\textbf{\textit{B}}}{\sqrt{\rho}}\, ,\label{paral}\\
{\boldsymbol\nabla\boldsymbol\cdot}\textbf{\textit{v}} & =& 0\, ,\label{konti2}\\
{\boldsymbol\nabla\boldsymbol\cdot}\textbf{\textit{B}} & = & 0\label{divb} \, .
\end{eqnarray}
The combination of Equation (\ref{konti2}) and Equation (\ref{paral}) yields
the conserved values, $\textbf{\textit{B}}{\boldsymbol\cdot\boldsymbol\nabla}\rho=0$ and
$\textbf{\textit{B}}{\boldsymbol\cdot\boldsymbol\nabla} M_{A}=0$, and therefore also
$\textbf{\textit{v}}{\boldsymbol\cdot\boldsymbol\nabla}\rho=0$ and $\textbf{\textit{v}}{\boldsymbol\cdot\boldsymbol\nabla} M_{A}=0$.
Consequently, the Alfv\'en Mach number and the density are constant along field lines and
the magnetic and the velocity field are integrable, i.e. non-ergodic \citep{GradRubin1958, 1970AmJPh..38..494S}.

Integrable, divergence-free fields, such as the magnetic field, can be represented by so 
called Euler or Clebsch potentials, e.g., $f$ and $g$, via the form
\begin{equation}
\textbf{\textit{B}}={\boldsymbol\nabla} f{\boldsymbol\times}{\boldsymbol\nabla} g\, .
\label{euler00}
\end{equation}
In general, these Euler potentials are functions of all three coordinates $x, y, z$.
The representation can also be made by alternative Euler potentials, say $\alpha$ and $\beta$,
if these are related to the original ones via the mapping $\alpha=\alpha(f,g)$ and 
$\beta=\beta(f,g)$, and if, in addition, the Poisson bracket is identical to unity, 
meaning that 
\begin{eqnarray}
\left[ \alpha,\beta\right]_{f,g}:=\frac{\partial\alpha}{\partial f}\frac{\partial\beta}
{\partial g}
-\frac{\partial\alpha}{\partial g}\frac{\partial\beta}{\partial f}\equiv 1 \, .
\label{poisson1}
\end{eqnarray}
Then the field remains unchanged and can be written as
\begin{equation}
\textbf{\textit{B}}={\boldsymbol\nabla} f{\boldsymbol\times}{\boldsymbol\nabla} g={\boldsymbol\nabla}\alpha{\boldsymbol\times}{\boldsymbol\nabla}\beta
=\left[ \alpha,\beta\right]_{f,g}\,{\boldsymbol\nabla} f{\boldsymbol\times}{\boldsymbol\nabla} g\, .
\label{euler1}
\end{equation}
This kind of transformation is called canonical transformation.

A non-canonical, hence \lq active\rq~transformation, on the other hand, is 
performed in case the Poisson bracket is not identical to unity.  
It was shown by \citet{1992PhFlB...4.1689G}  that such an active transformation
reflects the similarity between MHS states and stationary states in incompressible
MHD. This can be seen from the following.

If we start from the momentum equation of MHS given by 
\begin{equation}
{\boldsymbol\nabla} p_{S}=\textbf{\textit{j}}_{S} {\boldsymbol\times}\textbf{\textit{B}}_{S} = \left({\boldsymbol\nabla}{\boldsymbol\times}\textbf{\textit{B}}_{S}\right){\boldsymbol\times}\textbf{\textit{B}}_{S}
\label{MHS0}
\end{equation}
and represent the MHS magnetic field, $B_{S}$, via the Euler potentials $f$ and $g$ 
(where in the following the Euler potentials $f$ and $g$ refer to MHS fields and 
$\alpha$ and $\beta$ to stationary MHD fields)
\begin{equation}
\textbf{\textit{B}}_{S}  =  {\boldsymbol\nabla} f{\boldsymbol\times}{\boldsymbol\nabla} g\, , \label{defmhs}
\end{equation}
then the MHS pressure, $p_{S}$, can always be written locally as an explicit function of 
$f$ and $g$
\begin{equation}
p_{S}  =  p_{S}(f,g)\, . \label{Druck}
\end{equation} 
Let us now assume we know a solution ($p_{S}, \textbf{\textit{B}}_{S}$) for Equation\,(\ref{MHS0}) in which the magnetic field and the pressure
are given in the form of the Eqs.\,(\ref{defmhs}) and (\ref{Druck}).
If we additionally define a relation between the  Alfv\'{e}n Mach number $M_{A}$ and 
the Poisson bracket of the form 
\begin{equation}
\left(\left[ f,g \right]_{\alpha,\beta}\right)^2  \equiv  1-M_{A}^2  >  0\\
\end{equation} 
or, equivalently,
\begin{equation}
\left(\left[ \alpha,\beta\right]_{f,g}\right)^2  \equiv  \frac{1}{1-M_{A}^2} \geq 1 \label{poisson2}
\end{equation} 
where $M_{A}$ can always be regarded as an explicit function of $\alpha$  and
$\beta$ (or $f$ and $g$) bounded by one, then 
\begin{equation}
\textbf{\textit{B}}  =  {\boldsymbol\nabla} \alpha{\boldsymbol\times}{\boldsymbol\nabla} \beta\, 
\end{equation}
can be considered as magnetic field of a stationary MHD equilibrium.
This means that the corresponding velocity field can be written as
%
\begin{equation}
\textbf{\textit{v}} = \frac{M_{A}\textbf{\textit{B}}}{\sqrt{\rho}}\, ,\label{paral2}
\end{equation}
while the magnetic field, the corresponding current density, and the plasma pressure take the form
\begin{eqnarray}
\textbf{\textit{B}} &=&\left[ \alpha,\beta \right]_{f,g} \textbf{\textit{B}}_{S}
\equiv \frac{\textbf{\textit{B}}_{S}}{\sqrt{1-M_{A}^2}}\, , \label{stationaryB}\\
\textbf{\textit{j}} &=&\frac{M_{A}\,{\boldsymbol\nabla} M_{A}{\boldsymbol\times}\textbf{\textit{B}}_{S}}{\left(1-M_{A}^2\right)^{3/2}} 
+\frac{\textbf{\textit{j}}_{S}}{\left(1-M_{A}^2\right)^{1/2}}\, , \label{currenttrafo}\\
p &=& p_{S}-\frac{1}{2} \rho \textbf{\textit{v}}^2 \label{druck}
\end{eqnarray}
for sub-Alfv\'enic flows, and 
\begin{eqnarray}
\textbf{\textit{B}} &=&\left[ \alpha,\beta \right]_{f,g} \textbf{\textit{B}}_{S}
\equiv \frac{\textbf{\textit{B}}_{S}}{\sqrt{M_{A}^2-1}}\, , \\
\textbf{\textit{j}} &=&-\frac{M_{A}\,{\boldsymbol\nabla} M_{A}{\boldsymbol\times}\textbf{\textit{B}}_{S}}{\left(M_{A}^2-1\right)^{3/2}}
+\frac{\textbf{\textit{j}}_{S}}{\left(M_{A}^2-1\right)^{1/2}}\, ,\\
p &=& p_{0}-p_{S}-\frac{1}{2} \rho \textbf{\textit{v}}^2
\end{eqnarray}
for super-Alfv\'enic flows, implying that the  stationary MHD equations 
(Eqs.(\ref{konti})--(\ref{divb})) are fulfilled.
The parameter $p_{0}$ represents hereby a pressure offset, necessary to avoid negative 
pressure values and to provide boundary conditions.
In any case, the plasma density, $\rho$, and the 
Alfv\'enic Mach number are explicit functions of the Euler potentials $f$ and $g$.
If these can be constrained by reasonable boundary conditions (e.g. from observations),
the velocity and pressure, and correspondingly the complete stationary equilibrium, 
can be calculated from a known solution of $p_{S}$ and $\textbf{\textit{B}}_{S}$. One property of the 
transformation is that the geometrical and topological
field-line structures of the initial MHS state remain unchanged. 
A second one is that the flow induces current fragmentation whereby the flow itself is generated
via variations of the pressure. Current fragmentation induced by pressure pulses that
originate close to magnetic null points were also reported by \citet{2015ApJ...812..105J}.

\subsection{General parametrization of the transformation}

In the previous section we showed that a transformation method exists. What is 
needed next is to find a way to calculate explicitly the transformation
from the initial potentials $f$ and $g$ to the final ones $\alpha$ and $\beta$. 

The sub-Alfv\'enic Poisson bracket relation Equation\,(\ref{poisson2}) and, therefore, also 
the sub-Alfv\'enic $M_{A}$ can generally be represented via
\begin{eqnarray}
 M_{A}  &\equiv & \tanh {\cal M}(f,g)\nonumber\\
\!\!\!\!\!\!\!\land\,\,\,\left(\left[ \alpha,\beta\right]_{f,g}\right)^2  &\equiv &  \frac{1}{1-M_{A}^2}\equiv \left(\cosh {\cal M}(f,g)\right)^2
\geq 1 .
\label{bound1}\end{eqnarray}
The function $\cal M$ should be at least twice continuously differentiable. The condition
Equation\,(\ref{bound1}) guarantees that the Alfv\'en mach number is bounded by one. 
Keeping the polarity of the mapped magnetic field (see Equation\,(\ref{euler1})), Equation\,(\ref{bound1}) results in a linear 
partial differential equation for $\alpha$ and $\beta$ as functions of $f$ and $g$
\begin{eqnarray}
\left[ \alpha,\beta\right]_{f,g}:=\frac{\partial\alpha}{\partial f}\frac{\partial\beta}
{\partial g}
-\frac{\partial\alpha}{\partial g}\frac{\partial\beta}{\partial f}
\equiv \cosh {\cal M }(f,g),
\label{bound2}\end{eqnarray}
which could basically be solved based on the method of characteristics.

Searching for a method to reduce Equation\,(\ref{bound2}) to
a generally simpler form, can be done by assuming without loss of generality
\begin{eqnarray}
\alpha_{0} &=& \alpha_{0} (f)
\\
\nonumber\\
\beta_{0} &=& \left(\frac{d\alpha_{0}}{df} \right)^{-1} \int \cosh {\cal M}\left(f,g\right)\, dg
+ \beta_{00}\left(f\right)
\end{eqnarray}
which automatically satisfies Equation\,(\ref{bound2}). The functions $\alpha_{0}(f)$ and
$\beta_{00}(f)$ can be chosen arbitrarily to satisfy boundary conditions and 
constraints for the magnetic and the velocity fields. All equivalent
transformations $\alpha=\alpha(f,g)$ and $\beta=\beta(f,g)$ can be found by corresponding canonical
transformations of $\alpha_{0}$ and $\beta_{0}$.

\subsection{Basic equations for 2D and 2.5D MHS equilibria}

The general solution for stationary equilibria presented in the previous section 
is valid in all dimensions. Ideally, 3D stationary equilibria would be desired. 
To compute such equilibria via the transformation method requires the knowledge of exact 
and analytical 3D MHS equilibria. However, only few such 3D MHS equilibria are known
\citep{1991ApJ...370..427L, 1995A&A...301..628N, 1997A&A...325..847N, 1999GApFD..91..269P}. 
Nevertheless, for many practical scenarios the field geometry displays some symmetry. 
Translationally invariant equilibria serve as examples. These can be associated, e.g., 
with arcade structures above the polarity inversion line (PIL). These PILs resemble 
the $z$-axis (here: the invariant direction) in the topological sense. Therefore, a 
2D or 2.5D (which means that $B_{z}$ is nonzero) treatment is reasonable and provides 
a sufficiently 
accurate approximation with respect to the physical insights. The advantage of 2D and 
2.5D equilibria is that a wide-spread number of classes of magnetic configurations can 
be computed based on the well-known Grad-Shafranov (or, equivalently, L\"ust-Schl\"uter) 
theory \citep[see][]{1958JETP....6..545S, 1957ZNatA..12..850L}. According to this theory, 
one needs to solve the equilibrium condition
\begin{equation}
\Delta A = -\frac{d}{dA} \left( p_{S} + \frac{B_{z S}^{2}}{2}\right) = 
-\frac{d\Pi_{S}}{dA}\, ,\label{GSLS}
\end{equation}
which follows from the assumption of translational invariance ($\partial/\partial z \equiv 0$)
and the representation of the magnetic field by 
\begin{equation}
\textbf{\textit{B}}_{S} = {\boldsymbol\nabla} A(x,y){\boldsymbol\times} {\boldsymbol\nabla} z + B_{z S}(x,y)\textbf{\textit{e}}_{z}\, . \label{BS} 
\end{equation}
$B_{z S}$ is the so-called toroidal component \citep[see, 
e.g.,][]{1978mfge.book.....M,2006pspa.book.....S}.
$B_{z S}$ and $p_{S}$ are necessarily explicit functions of the flux function $A$. 
To solve Equation\,(\ref{GSLS}) a physically motivated pressure function $\Pi_{S}$ has to be defined.  

Solutions of the Equation\,(\ref{GSLS}) are solutions to the MHS equations
\begin{eqnarray}
{\boldsymbol\nabla} p_{S} & = & \left({\boldsymbol\nabla}{\boldsymbol\times}\textbf{\textit{B}}_{S}\right){\boldsymbol\times}\textbf{\textit{B}}_{S}\, ,\label{MHS1}\\
{\boldsymbol\nabla\boldsymbol\cdot} \textbf{\textit{B}}_{S} & = & 0\, . \label{MHS2}
\end{eqnarray}
The two systems of equations (Eqs.\,(\ref{GSLS})-(\ref{BS}) and Eqs.\,(\ref{MHS1})-(\ref{MHS2}))
are equivalent. 

The strategy is hence the following: We first need to solve the static Grad-Shafranov equation
to obtain an MHS equilibrium suitable to describe solar arcade structures. Then, a reasonable
mapping needs to be found that transforms this MHS equilibrium into a stationary state.

\section{Results}
\subsection{Mapping from 2D to current sheets varying in $z$-direction}
\label{sect.2D}

First we want to show that even pure 2D fields can be mapped to stationary fields which 
depend also on the $z$-direction. A translational invariant magnetic field can be written as 
$\textbf{\textit{B}}_{S}={\boldsymbol\nabla} A(x,y){\boldsymbol\times}{\boldsymbol\nabla} z$ in which the flux function $A$ depends only on
$x$ and $y$, and the electric current has only a $z$-component, as is obvious from 
$\textbf{\textit{j}}_{S} \equiv -\Delta A \textbf{\textit{e}}_{z}$.
Comparison with the definition of the static magnetic field (Equation\,(\ref{defmhs})) then
implies that $f$ must be identical to $A(x,y)$ and $g$ to $z$.
With this definition of the magnetic field, 
the Grad-Shafranov equation that needs to be solved reduces to
\begin{equation}
\Delta A = -\frac{dp_{S}}{dA}\, .
\end{equation}

For the transformation to the stationary magnetic field (Equation\,(\ref{stationaryB})), the Poisson 
bracket has to be evaluated. This is done in the following way 
\begin{eqnarray}
\left[\alpha,\beta \right]_{f,g} & = & \frac{\partial\alpha}{\partial f}\frac{\partial\beta}
{\partial g}
-\frac{\partial\alpha}{\partial g}\frac{\partial\beta}{\partial f} \\
 &= & \frac{\partial\alpha(A,z)}{\partial A}\frac{\partial\beta(A,z)}
{\partial z} -\frac{\partial\alpha(A,z)}{\partial z}\frac{\partial\beta(A,z)}{\partial A}\, . \nonumber
\end{eqnarray}
The dependence of the Poisson bracket, and therefore of the Mach number, 
on $z$ implies that the application of a non-canonical transformation to translational invariant
MHS-equilibria creates a magnetic field and a velocity field which can vary in the former
invariant direction. From inspection of Equation\,(\ref{stationaryB}) it is obvious that 
the geometry of the field lines (and therefore their direction) remains unchanged, while
the amplitude of the transformed fields is different from the original one and varies non-linearly 
with $z$.

By exploiting that $M_{A}$ is an explicit function of the
static Euler potentials $A$ and $z$, the electric current of the transformed field can be 
evaluated via the relation Equation\,(\ref{currenttrafo}). It results to 
\begin{eqnarray}
\textbf{\textit{j}} &=& \frac{M_{A}\frac{\partial M_{A}}{\partial z}\,{\boldsymbol\nabla} A -\textbf{\textit{e}}_{z}\left(
M_{A}\frac{\partial M_{A}}{\partial A}\left({\boldsymbol\nabla} A\right)^2\right)}
{\left(1-M_{A}^2\right)^{3/2}} \nonumber \\
 & & -\frac{\Delta A\, \textbf{\textit{e}}_{z}}{\left(1-M_{A}^2\right)^{1/2}}\, . 
\end{eqnarray}
As $A$ is a function of $x$ and $y$, it is obvious that the electric current of the 
transformed field has now components in all three coordinate directions which also depend
non-trivially and non-linearly on all three coordinates. It is hence
quasi-3D, but the field line structure in each $x,y$-plane is preserved. 
These additional current components, which are all perpendicular
to the magnetic field, guarantee self-consistently
that the system is kept in equilibrium state. Moreover, the current density deviates
from the one of the pure 2D MHS field, which has only a current component in $z$-direction.
Hence, despite the fact that we started from an initially highly symmetric configuration, 
the resulting current displays a much more complex structure.

\subsection{Mapping from 2.5D to 3D}\label{sect:map2p5}

The magnetic field of solar arcade structures must not necessarily consist of field lines
that lie purely in ($x,y$)-planes laminated in $z$-direction. Instead, the field lines could 
possess a helical structure, which means that the magnetic field has a toroidal component pointing 
in $z$-direction. Such cases require at least a 2.5D treatment. We refrain here from discussing 
full 3D scenarios, because they cannot be solved using the Grad-Shafranov theory anymore.

To compute 2.5D MHS equilibria, we need to solve the full Grad-Shafranov equation (\ref{GSLS}).
The representation of the MHS field via Euler potentials is more tricky
in the 2.5D case, because at least one of the Euler potentials has to depend on all
three spatial coordinates and must depend linearly on $z$. Hence, we need to construct
such an Euler potential.

The simplest case would be to keep for $f$ the same prescription as in the 2D case, i.e., 
$f=A(x,y)$, and to assume that $g$ can be defined as $g=z+\tilde{h}(x,y)$.
The function $\tilde{h}(x,y)$ can be chosen such that at least locally, it can be expressed
by the flux function $A$ via $\tilde{h}(x,y) = h(A(x,y),y)$.  Such a choice of representation
is motivated by the fact that $A$ has the strongest variation in $x$-direction if the coordinate
system is chosen in such a way that the $y$-direction corresponds to the vertical axis 
of the arcade structures, i.e., it is perpendicular to the solar surface.

While usually the Euler potentials are used to compute the $B_{zS}$ component 
\citep[e.g.][]{2006pspa.book.....S}, this cannot
be done so easily anymore for the current representation of the Euler potentials, because
the function $h$ is not known. Therefore, one needs first to evaluate $B_{zS}$ from
the Grad-Shafranov equation (\ref{GSLS}), and only then the function $h$ can be 
determined under some constraints. When comparing the Euler representation for the magnetic
field with the representation via the Grad-Shafranov equation 
\begin{equation}
\textbf{\textit{B}}_{S} ={\boldsymbol\nabla} f{\boldsymbol\times}{\boldsymbol\nabla} g\equiv {\boldsymbol\nabla} A{\boldsymbol\times}{\boldsymbol\nabla} z+ 
B_{zS}\,\textbf{\textit{e}}_{z}\, ,
\end{equation}
it follows that
\begin{equation}
 {\boldsymbol\nabla} A{\boldsymbol\times}{\boldsymbol\nabla} h = {\boldsymbol\nabla} A{\boldsymbol\times} \frac{\partial h}{\partial y}\,\textbf{\textit{e}}_{y}\equiv B_{zS}(A)\,\textbf{\textit{e}}_{z}\, .\label{id1}
\end{equation}
Scalar multiplication of the identity Equation\,(\ref{id1}) with $\textbf{\textit{e}}_{z}$ leads to
\begin{equation}
\frac{\partial h}{\partial y}\bigg|_{A}\,\frac{\partial A}{\partial x}\bigg|_{y} = - \frac{\partial h}{\partial 
y}\,B_{yS}(A,y) = B_{zS}(A)\, , \label{pde1}
\end{equation}
where $\frac{\partial A}{\partial x}=-B_{yS}(A,y)$ has to be considered as a
function of the chosen coordinates $A$ and $y$, because the partial differential equation Equation\,(\ref{pde1})
for $h$ has a solution, which is a function of these coordinates.

The function $h(A,y)$ can thus be computed from
\begin{eqnarray}
h(A,y) & = & - \int\,\frac{B_{zS}(A)}{B_{yS}(A,y)}\, dy + h_{0}(A) \nonumber \\
& = & - B_{zS}(A)\int\,\frac{dy}{B_{yS}(A,y)}\, + h_{0}(A)\, .
\label{shearpot1}
\end{eqnarray}
One should note, however, that the evaluation of the function $h(A,y)$ bears difficulties,
for example, if the magnetic field has null points. In that case, $B_{xS}(A,y) = B_{yS}(A,y) = 0$
and $\frac{\partial h}{\partial y}$ diverges. Therefore, to properly define a function $h(A,y)$ 
in the vicinity of a null point, the toroidal component $B_{zS}(A)$ 
must be zero on the separatrix surface, i.e., $B_{zS}
(A_{sep})=0$ with ${\boldsymbol\nabla} A|_{x_{N},y_{N}}=0$, if the null point is of X-point type. In case of 
an O-point null point, $B_{zS}(A)$ has to vanish at that point.    

To perform the transformation we recall that the Alfv\'en Mach number $M_{A}$ is an explicit 
function of the static Euler potentials $f=A(x,y)$ and $g=z+h(A(x,y),y)$. The relations 
Eqs.\,(\ref{pde1})--(\ref{shearpot1}) provide a representation of these Euler 
potentials and, therefore, the basis for the definition of $M_{A}$. 
Hence, the electric current of the
transformed field can be evaluated via the relation Equation\,(\ref{currenttrafo}). It results to

%
\begin{eqnarray}
\textbf{\textit{j}} & =&M_{A}\,\frac{\left(\frac{\partial M_{A}}{\partial A}{\boldsymbol\nabla} A +
\frac{\partial M_{A}}{\partial g}{\boldsymbol\nabla} g\right){\boldsymbol\times}
\left({\boldsymbol\nabla} A{\boldsymbol\times}{\boldsymbol\nabla} z+ B_{zS}\,\textbf{\textit{e}}_{z}\right)}
{\left(1-M_{A}^2\right)^{3/2}}\nonumber\\
&& +\frac{-\Delta A\, \textbf{\textit{e}}_{z}+B_{zs}'(A){\boldsymbol\nabla} A{\boldsymbol\times}\textbf{\textit{e}}_{z}}
{\left(1-M_{A}^2\right)^{1/2}}
\\
& =& M_{A}\,\frac{
\left(\frac{\partial M_{A}}{\partial g}
\right)\,{\boldsymbol\nabla} A - \textbf{\textit{e}}_{z}\left(
\frac{\partial M_{A}}{\partial A}\left({\boldsymbol\nabla} A\right)^2
+\frac{\partial M_{A}}{\partial g}{\boldsymbol\nabla} A{\boldsymbol\cdot\boldsymbol\nabla} h\right)}
{\left(1-M_{A}^2\right)^{3/2}}\nonumber\\
&& + M_{A}\,\frac{\frac{\partial M_{A}}{\partial A} B_{zS} {\boldsymbol\nabla} A{\boldsymbol\times}\textbf{\textit{e}}_{z}
+\frac{\partial M_{A}}{\partial g} B_{zS} {\boldsymbol\nabla} h{\boldsymbol\times}\textbf{\textit{e}}_{z}}
{\left(1-M_{A}^2\right)^{3/2}}\nonumber\\
\nonumber\\
&& +\frac{-\Delta A\, \textbf{\textit{e}}_{z}+B_{zs}'(A){\boldsymbol\nabla} A{\boldsymbol\times}\textbf{\textit{e}}_{z}}{\left(1-M_{A}^2\right)^{1/2}}
\\
\nonumber\\
&=&  \frac{M_{A}}{\left(1-M_{A}^2\right)^{3/2}}\left[
\left(\frac{\partial M_{A}}{\partial g}
\right)\,{\boldsymbol\nabla} A \right. \nonumber\\
& & \left.- \textbf{\textit{e}}_{z}\left(\left(\frac{\partial M_{A}}{\partial A} + \frac{\partial M_{A}}{\partial g}
\frac{\partial h}{\partial A}
\right)\left({\boldsymbol\nabla} A\right)^2
+\frac{\partial M_{A}}{\partial g}\frac{\partial h}{\partial y}\frac{\partial A}{\partial y}
\right)\right.
\nonumber\\
&& + \left.
B_{zS}\left(\frac{\partial M_{A}}{\partial A}
+\frac{\partial M_{A}}{\partial g}\frac{\partial h}{\partial A}\right){\boldsymbol\nabla} A{\boldsymbol\times}\textbf{\textit{e}}_{z} \right.\nonumber\\
& &\left. +B_{zS}\frac{\partial M_{A}}{\partial g}\frac{\partial h}{\partial y}\textbf{\textit{e}}_{x}\right]
+\frac{-\Delta A\, \textbf{\textit{e}}_{z}+B_{zs}'(A){\boldsymbol\nabla} A{\boldsymbol\times}\textbf{\textit{e}}_{z}}{\left(1-M_{A}^2\right)^{1/2}}
\end{eqnarray}
\begin{eqnarray}
&=&  \frac{M_{A}}{\left(1-M_{A}^2\right)^{3/2}}\left[
\left(\frac{\partial M_{A}}{\partial g}
\right)\,{\boldsymbol\nabla} A \right.\nonumber\\
& & \left. - \textbf{\textit{e}}_{z}\left(
\left(\frac{\partial M_{A}}{\partial A} + \frac{\partial M_{A}}{\partial g}
\frac{\partial h}{\partial A}
\right)\left({\boldsymbol\nabla} A\right)^2
-\frac{\partial M_{A}}{\partial g} B_{zS}\frac{B_{xS}}{B_{yS}}
\right)\right.
\nonumber\\
&& + \left.
B_{zS}\left(\frac{\partial M_{A}}{\partial A}
+\frac{\partial M_{A}}{\partial g}\frac{\partial h}{\partial A}\right){\boldsymbol\nabla} A{\boldsymbol\times}\textbf{\textit{e}}_{z}\right.\nonumber\\
& &\left. - B_{zS}\frac{\partial M_{A}}{\partial g}\frac{B_{zS}}{B_{y}}\textbf{\textit{e}}_{x}\right]
+\frac{-\Delta A\, \textbf{\textit{e}}_{z}+B_{zs}'(A){\boldsymbol\nabla} A{\boldsymbol\times}\textbf{\textit{e}}_{z}}{\left(1-M_{A}^2\right)^{1/2}}
\end{eqnarray}
As before (Sect. \ref{sect.2D}), the variations of the current are induced by the flow, which itself is
generated by the non-canonical mapping. 

The most interesting result is the occurrence of a current component parallel to the poloidal 
magnetic field component. Such a component does not exist in a 2D mapping
of a pure poloidal field\footnote{For a translational invariant magnetic field only $x$-$y$ components
exist in the poloidal plane and only one \lq toroidal\rq~component of the current, namely in
$z$-direction, exist.} and also not in the quasi-laminar regime discussed in Sect.\,\ref{sect.2D}, where only 
an additional component in ${\boldsymbol\nabla} A$-direction exists due to the change of the Mach number
in $z$-direction. This additional poloidal component ${\boldsymbol\nabla} A{\boldsymbol\times} e_{z}$ of the current exists  
not only due to the static component
$B_{zS}'(A){\boldsymbol\nabla} A{\boldsymbol\times} e_{z}$, but 
due to the explicit dependence of $M_{A}$ on $A$ and $g$.
This latter is true even if the static component $B_{zS}$ is constant.

A current component into the main direction of the (poloidal) magnetic field
strengthens the character of the current towards a more field-aligned current.
Moreover, it provides the basis for particle acceleration, as a switched on resistivity
would generate an electric field with a strong component parallel to the magnetic field.

\subsection{(Non-)existence of 3D force-free fields}

The prerequisite of our investigations about stationary MHD flows and their current 
structures are MHS equilibria. Force-free states are an important sub-class of MHS states.
They correspond to states of minimum magnetic energy into which each 
equilibrium after distortion should relax according to variational calculus 
\citep[e.g.,][]{1979PASJ...31..209S}.

The following vivid illustration, that is based on the original ideas of 
\citet[][see page 126ff.]{1975MWGBI.........K} and \citet{1972ApJ...174..499P}, 
will help to elucidate why
force-free states occur. Let us consider that we have a 
small domain with an interlaced field topology, so that one field line is interwoven in such a 
way that this single field line fills basically the complete volume. Then, by knowing that 
the pressure is constant along each individual field line, 
\begin{equation}
{\boldsymbol\nabla} p_{S} = \left( {\boldsymbol\nabla}{\boldsymbol\times}\textbf{\textit{B}}_{S} \right) {\boldsymbol\times} \textbf{\textit{B}}_{S} \quad \Rightarrow
\quad \textbf{\textit{B}}_{S} {\boldsymbol\cdot} {\boldsymbol\nabla} p_{S} = 0\, , \label{force-free}
\end{equation}
which is a necessary MHS condition, it follows that the pressure is constant in this whole volume. 
This leads automatically to a force-free state, because 
${\boldsymbol\nabla} p_{S} = \left( {\boldsymbol\nabla}{\boldsymbol\times}\textbf{\textit{B}}_{S} \right) {\boldsymbol\times} \textbf{\textit{B}}_{S}  = \textbf{\textit{0}}$. 
In this context, a constant pressure inside the volume hence guarantees that influences from 
outside are switched off. 

In contrast, if the considered field line extends beyond the border of the domain, the condition
Equation\,(\ref{force-free}) implies that the pressure inside this volume is, at least partially, 
determined by constraints from outside \citep[see][]{1972ApJ...174..499P} and the state is
not necessarily force-free. 

Recent investigations, allowing at least for field deformations via boundary footpoint 
displacements, also minimize the influence on the outer boundaries of the MHS environment
e.g. by the severe assumption that the velocity of the footpoints should vanish at the 
boundary \citep[see, e.g.,][]{2010ApJ...718..717L, 2012ASSP...33....3P}. This means that no
flow can leave the volume, and any flows that might occur along field lines are basically ignored.

If we would be dealing with an exclusively magneto-hydrostatic atmosphere where
stationary flows could be completely excluded, the force-freeness could be a
reasonable assumption. However, as observations have shown, flows are naturally occurring
in the solar atmosphere \citep[e.g.,][]{1971PASJ...23..443Y, 1976SoPh...47..233H, 2010SoPh..267..361W} 
so that the MHS states are embedded in regions in which
locally flows can occur. Hence, it is not necessarily always possible to eliminate external 
influences, but, in contrast, the occurrence of flows can be utilized, because they 
help to determine exactly the \lq integral\rq~ parameters like the plasma pressure.

We want to test whether or under which circumstances the states after non-canonical transformations can be 
force-free. For this, we regard the transformation method in analogy to quantum-mechanics.
The set of equilibria (before and after the transformation) can be considered as a family of stationary states,
having all the identical (geometrically and topologically) field-line structure. In this family,  
the MHS equilibrium defines
the ground state ($M_{A} = 0$, for all $x, y, z$). All other stationary states with flow can then be 
regarded as excited states.  
With this interpretation in mind, there are two possible scenarios for which force-free states can be expected: 
(i) if the ground state is already force-free, meaning that $\textbf{\textit{j}}_{S}{\boldsymbol\times}\textbf{\textit{B}}_{S} = {\boldsymbol\nabla} p_{S} = 0$,
or (ii) if the original non-force free ground state turns into a force-free final state when performing the 
transformation, i.e., via the application of a flow.

In general the direction of the magnetic field remains unchanged under the transformation. If we demand
that the transformed field is force-free, the following equivalence is valid 
\begin{equation}
\textbf{\textit{j}}{\boldsymbol\times}\textbf{\textit{B}}=\textbf{\textit{0}}\Leftrightarrow \textbf{\textit{j}}{\boldsymbol\times}\textbf{\textit{B}}_{S}=\textbf{\textit{0}}\, . \label{equiv}
\end{equation} 
Inserting the general form of the transformed current (Equation\,(\ref{currenttrafo})), the equation on the
right-hand side of Equation\,(\ref{equiv}) delivers
\begin{equation}
\frac{\textbf{\textit{B}}_{S}^2 {\boldsymbol\nabla} M_{A}^2}{2\left(1-M_{A}^2\right)}  = {\boldsymbol\nabla} p_{S}\, . \label{ff1}
\end{equation} 

Let us start with the first case of a force-free ground state. Then Equation\,(\ref{ff1}) implies that
${\boldsymbol\nabla} p_{S} = 0$. This means that, without loss of generality, $M_{A}$ must be constant
throughout the whole considered domain. This is an extreme constraint for the whole non-linear
MHD flow and is only fulfilled in exceptionally rare cases.
A similar result was obtained by \citet{2005PhPl...12e2902K}, who investigated pure 2D nonlinear 
force-free magnetic fields with mass flow.
Field-aligned flows can be regarded as nonlinear perturbation
of the MHS state. In analogy to linear perturbation theory, i.e. linear 
stability analysis, we may say that any unstable mode that might occur
will occur. This means that if a self-consistent pressure 
perturbation\footnote{Here, self-consistent means that the pressure variation 
supports the field-aligned equilibrium flow.}, like the one given by 
Equation\,(\ref{druck}), will occur (not only at the footpoints of the magnetic field 
structure), the force-free magnetic field cannot be maintained. Therefore, we 
can conclude that in any region, in which nonlinear flows can occur and are not 
suppressed, force-free fields will not exist or will vanish.

Turning to the second case, we can decompose the pressure gradient ${\boldsymbol\nabla} p_{S} = {\boldsymbol\nabla} p_{S}(f,g)$
and the gradient of the square of the Alfv\'{e}n Mach number ${\boldsymbol\nabla} M_{A}^2$ (with $M_{A} = M_{A}(f,g)$)
in the following way: 
\begin{eqnarray}
\frac{\textbf{\textit{B}}_{S}^2}{2\left(1-M_{A}^2\right)}\frac{\partial M_{A}^2}{
\partial f} &&=\frac{\partial p_{S}}{\partial f}\, ,\label{bfrac1}\\
\frac{\textbf{\textit{B}}_{S}^2}{2\left(1-M_{A}^2\right)}\frac{\partial M_{A}^2}{
\partial g}&&=\frac{\partial p_{S}}{\partial g}\, .\label{bfrac2}
\end{eqnarray}
These two equations imply that the expression
$\textbf{\textit{B}}_{S}^2$ must be an explicit function of $f$ and $g$ only, and hence $\textbf{\textit{B}}_{S}^2$
as well as $\textbf{\textit{B}}^2$ must be constant along fieldlines. An additional
restriction is introduced by the fact that for a given MHS equilibrium (defined by $B_{S}, p_{S}$) 
two first-order differential equations 
result for {\it one} function, i.e. $M_{A}$. This implies that $M_{A}$ is an explicit function
of $\textbf{\textit{B}}_{S}^2$. Considering that $M_{A}^{2}$ represents the ratio between kinetic and magnetic energy 
density, this causes a strong correlation between the magnitude of the energy partition and the magnetic 
field strength. 
Such severe restrictions tremendously limit on the one hand the number of basic eligible MHS-equilibria 
that can be used for such a transformation, and on the other hand also the freedom of the choice of 
reasonable Mach numbers and consequently of the flow. This leads to the conclusion that force-free 
is not a generic result\footnote{The force-free paradigm for the solar corona 
plasma was also criticized by \citet{2015A&A...584A..68P} based on different 
physical aspects.} but will occur only for rare cases with severe constraints.

How can we interpret these findings? As we said earlier, the MHS solutions are 
in general a small subset of the field-aligned incompressible flows. 
As we could show, in almost every case, any of these flows
either destroys the initial force-free property of
the magnetic field, or the transformed equilibrium of 
arbitrary topology and geometry cannot be force-free anymore. 

The force-free property is not compatible with an equilibrium flow, having a larger
cardinality as the original set of MHS states. 
We excluded non-field-aligned flows in our investigation, as they do
not have this strong affinity to MHS states.

\begin{figure}[h]
\begin{center}
\includegraphics[width=0.41\textwidth]{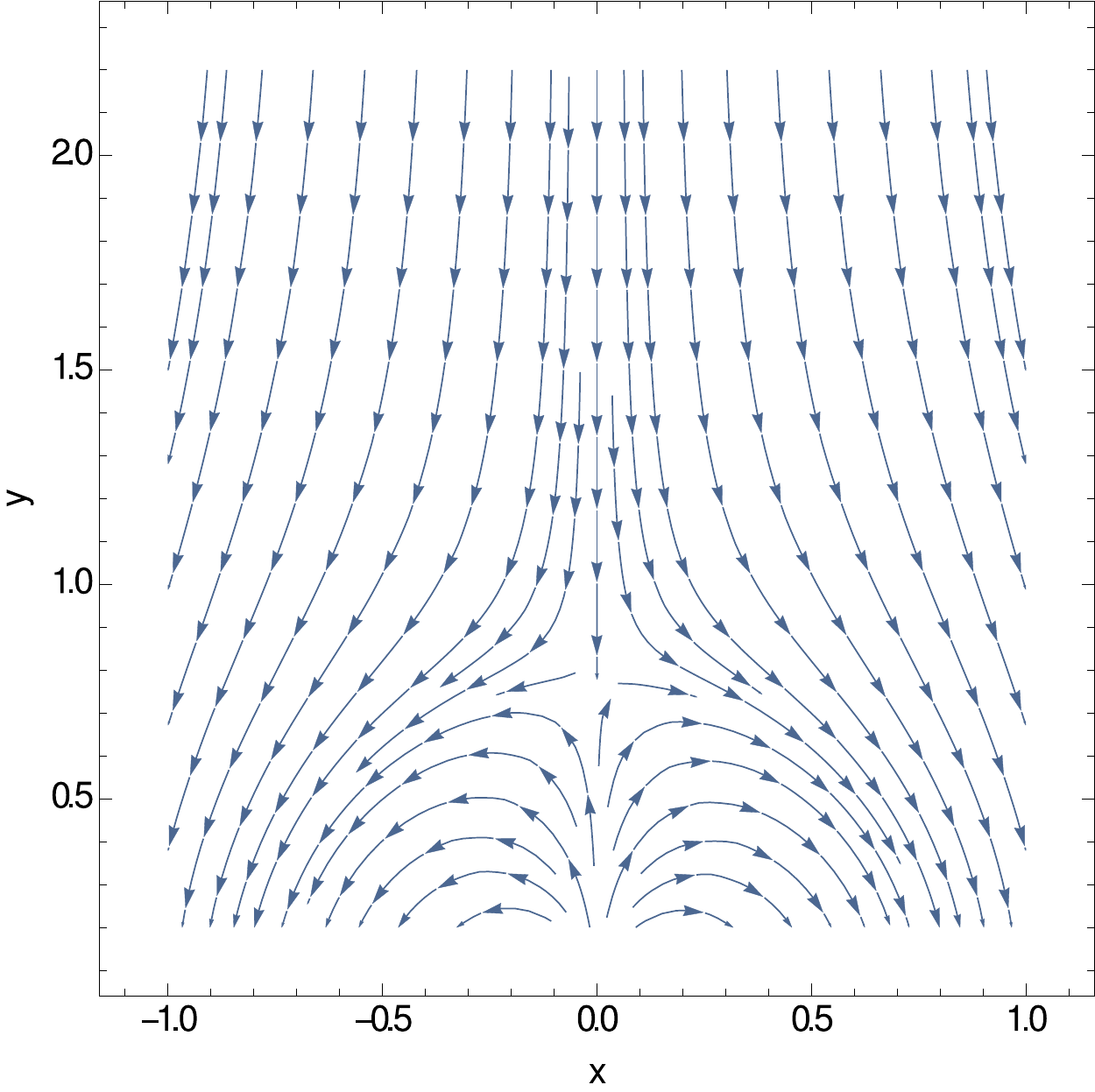}
\caption{2D potential field zoomed-in to the dipole region.}
\label{fig:feldli1}
\end{center}
\end{figure}

\section{Examples}

We wish to stress that the equations for the transformation of the current derived in Sects.\,\ref{sect.2D} and \ref{sect:map2p5} 
are generally valid and limited to neither a particular initial physical scenario nor to a specific flow pattern, determined by $M_{A}$.
In the following, for pure demonstration purpose, we chose two specific ground states, i.e. MHS states, one for a 2D equilibrium and 
the other for a 2.5D equilibrium. To each specific flow patterns are applied and the transformed current is 
computed. The Mach number profiles are chosen such that significant current fragmentation is 
achieved. Current fragmentation is an indispensable
physical process for plasma heating applications, e.g., in the solar corona. We thus pick a physical environment for our model calculations
which can be considered representative for (subareas) of coronal arcade structures and loops.

\subsection{2D scenario}

We start from a 2D potential field as a current free MHS state. To simulate the footpoint region of a typical solar arcade or
of some other mono-polar domain of the magnetic field in the solar atmosphere we superimpose a line-dipole, which is located at
the solar surface, and a homogeneous field. Our coordinate system has its origin on the solar surface with the $x$- and $z$-axis
being tangential to the surface and the $y$-axis perpendicular.

\begin{figure}[h]
\includegraphics[width=0.41\textwidth]{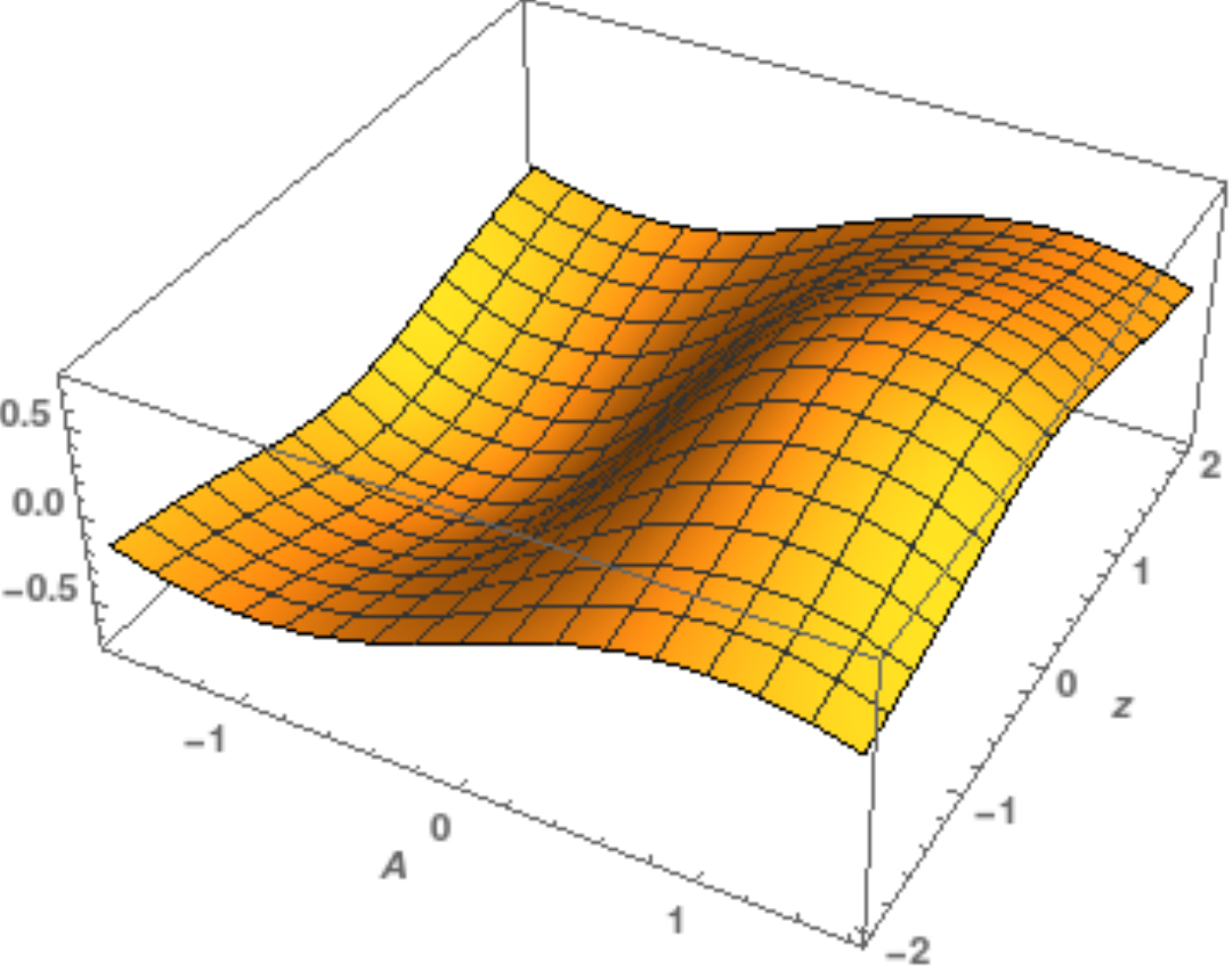}
\includegraphics[width=0.41\textwidth]{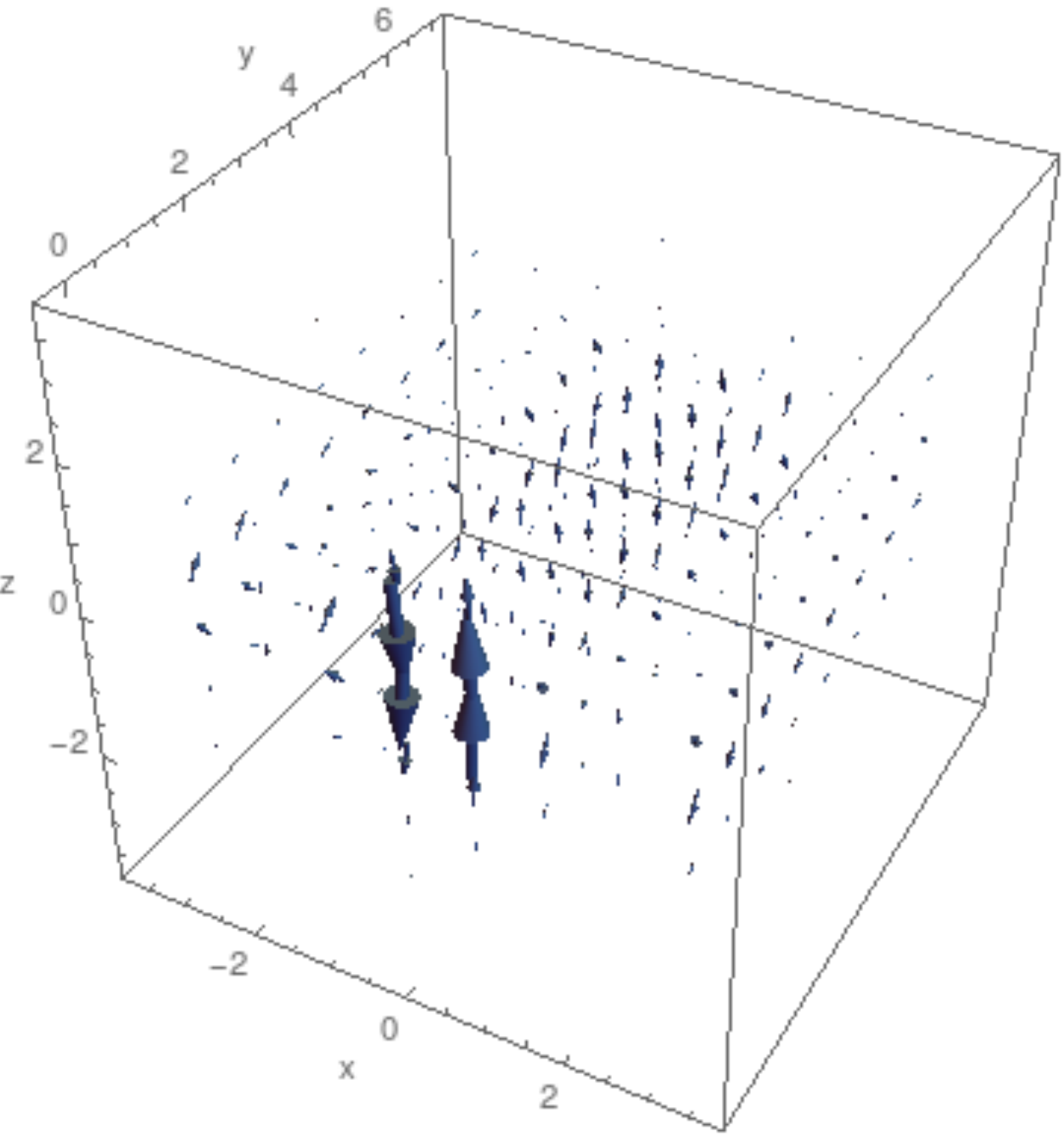}
\includegraphics[width=0.44\textwidth]{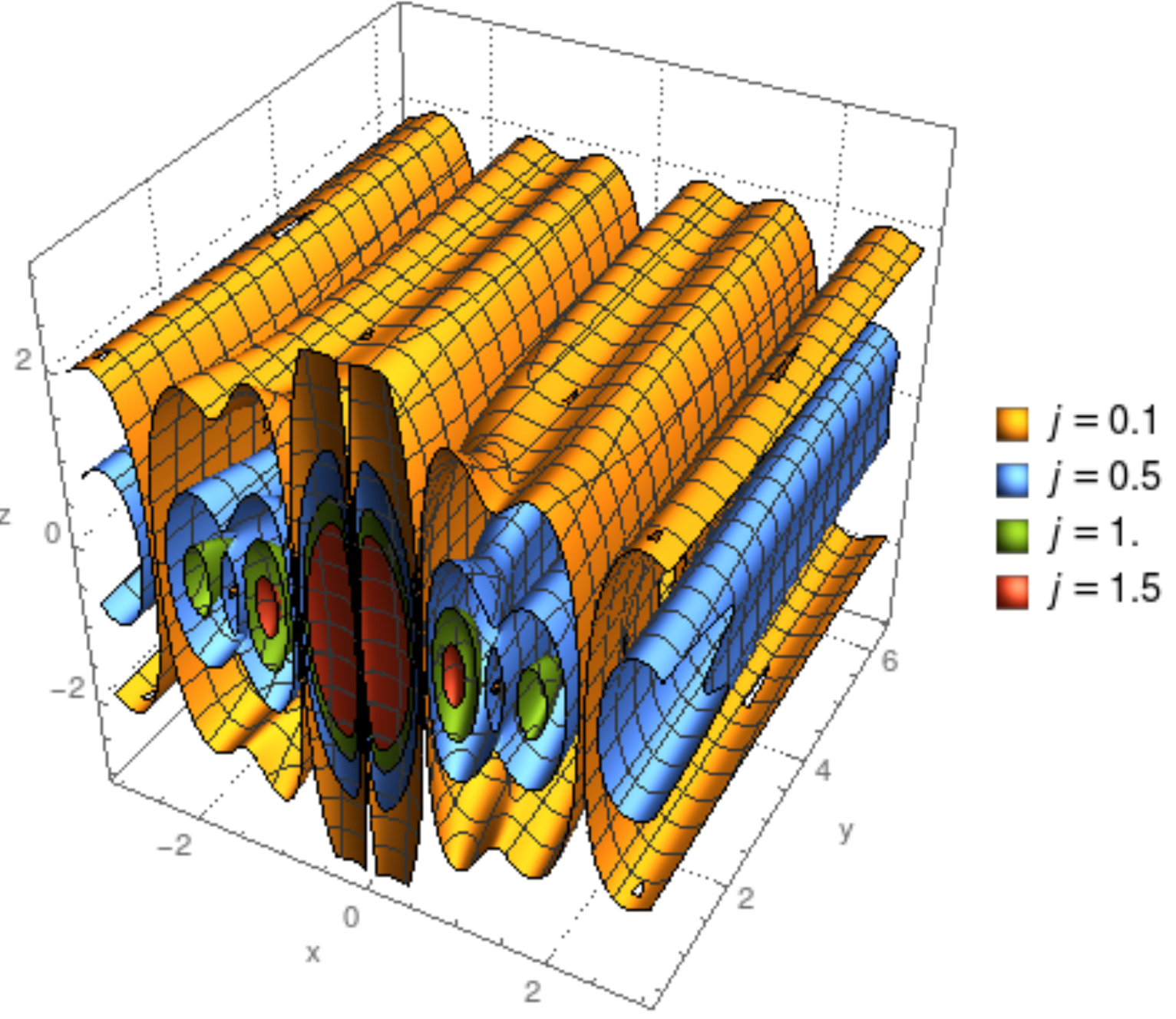}
\caption{Mach number profile (top) for the values of $k_1=1.57$ and $k_2=0$ plotted over the phase space $f\equiv A(x,y)$ and $g\equiv z$, current density vectors (mid) and isocontours of the current (bottom) for 
this Mach number profile}
\label{fig:cfrag1}
\end{figure}

The field configuration is computed from the complex potential     
${\cal A} \left(u \right)$ with  $u=x+iy$ and
\begin{equation}
{\cal A} \left(u \right)=i u - \frac{0.6 i}{u}\, ,
\end{equation}
where the imaginary part of ${\cal A}$ is the magnetic
flux function $A$
\begin{equation}
\Im({\cal A})= A(x,y)=x-\frac{0.6 x}{x^2+y^2}\, .
\end{equation}
The magnetic field then results to
\begin{eqnarray}
B_{xS}&=&\frac{\partial A}{\partial y}=-\frac{1.2 x y}{\left(x^2+y^2\right)^2}
\\
B_{yS}&=&-\frac{\partial A}{\partial x}=-1+\frac{0.6 y^2 - 0.6
x^2}{\left(x^2+y^2\right)^2}\, .
\end{eqnarray}
This process delivers an X-type magnetic null point at $(x_N,y_N)=(0,\sqrt{3/5})$
in the upper half domain $y>0$. For illustration, the field lines of this particular potential field are 
shown in Figure\,\ref{fig:feldli1}.

\begin{figure}[h]
\includegraphics[width=0.42\textwidth]{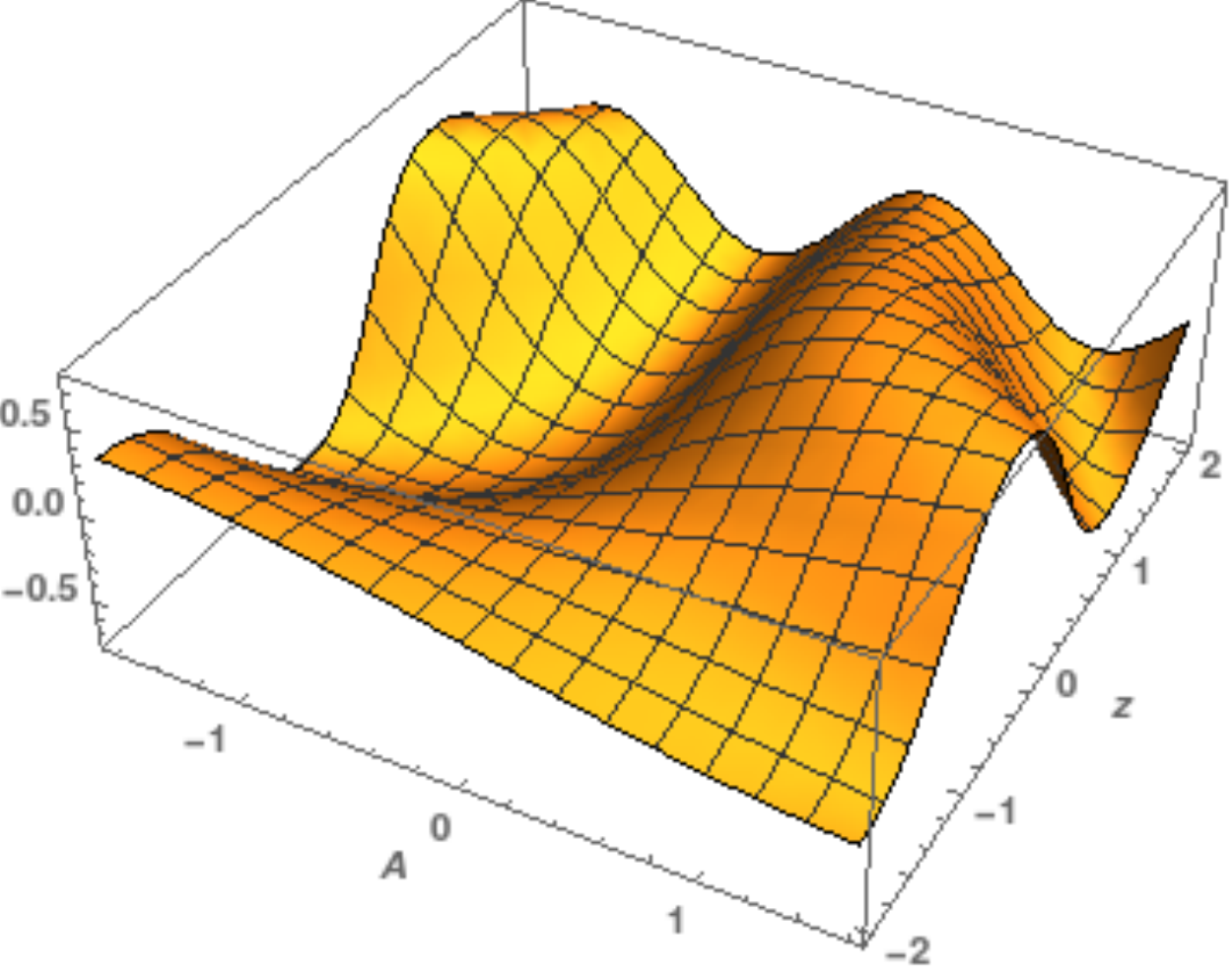}
\includegraphics[width=0.40\textwidth]{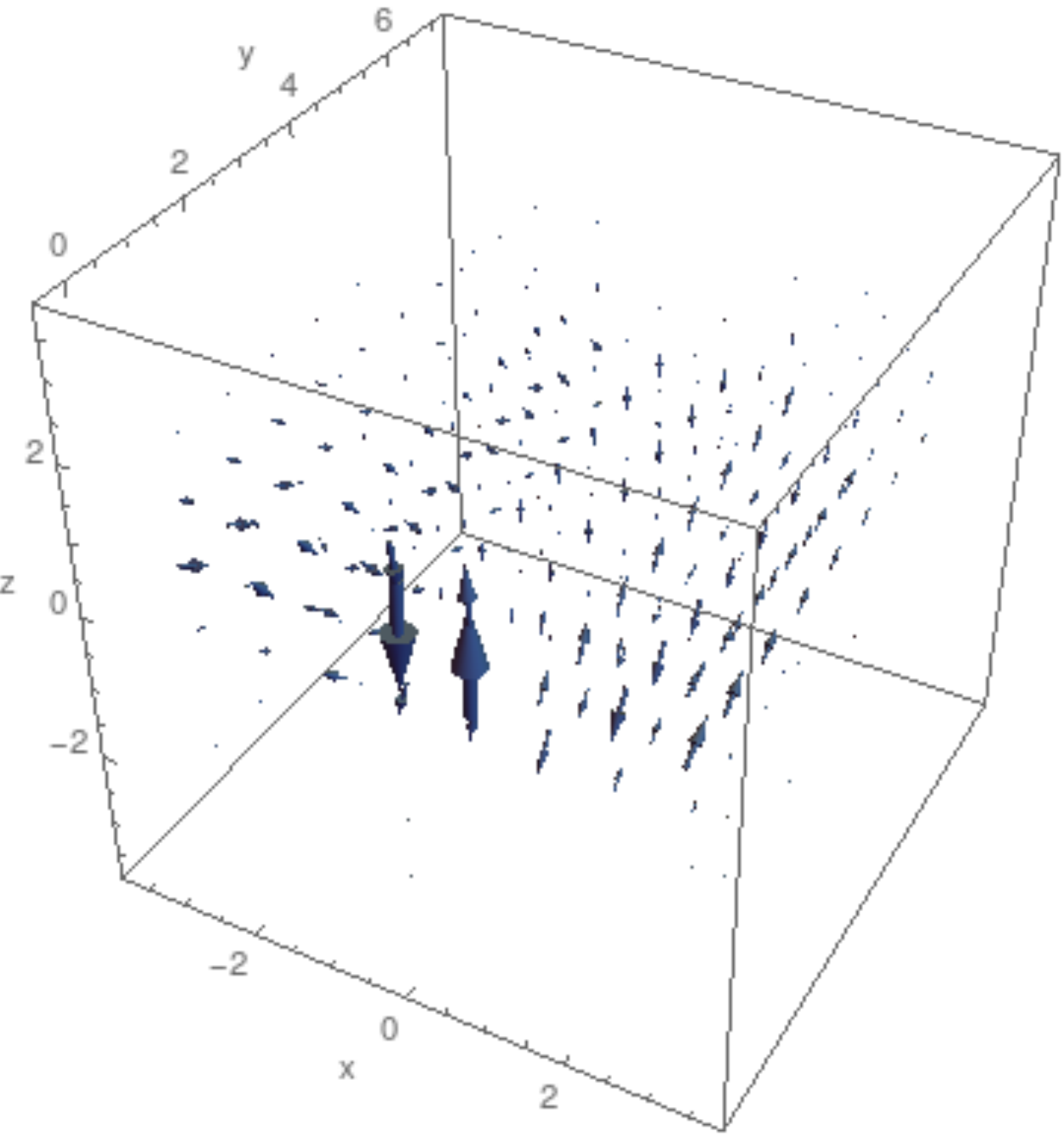}
\includegraphics[width=0.44\textwidth]{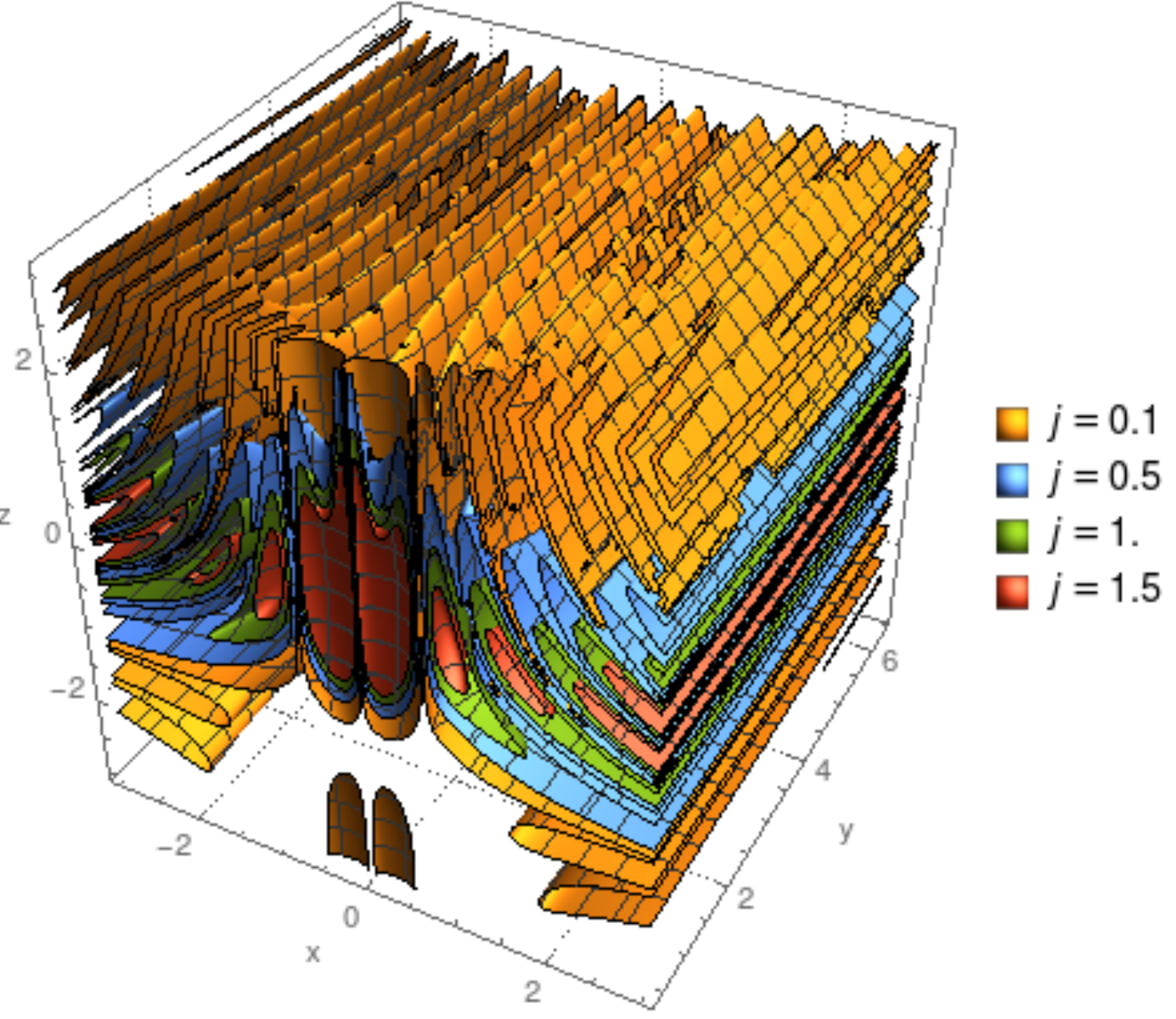}
\caption{Mach number profile (top) for the values of $k_1=1.57$ and $k_2=0.75$ plotted over the phase space $f\equiv A(x,y)$ and $g\equiv z$, current density vectors (mid) and isocontours of the current (bottom) for this Mach number profile.}
\label{fig:cfrag2}
\end{figure}

The general form of the Mach number profile is given by Equation\,(\ref{bound1}). We chose 
the function ${\cal M} (f,g) \equiv {\cal M} (A,z)$ in the following parametrized form
\begin{equation}
{\cal M} (A,z) = \left[\frac{\sin(k_1 A (1+k_2 z))}{1+ 0.5 z^{2}}\right]\, , 
\label{Mach_param}
\end{equation}
where the parameters $k_1$ and $k_2$ are constants. The choice of the sine function guarantees that the Mach number profile 
has a wavy shape, which causes gradients that produce great spacial variations in the resulting current.
A non-constant Mach number that varies spatially on small scales is motivated by the analogy to perturbation theory.
Every flow induced by the Mach number should optimize the current distribution to guarantee efficient dissipation of magnetic 
energy in form of Ohmic heating.

To study the influence of the choice of the Mach number profile on the resulting current structure, we compute two scenarios, one
of them is symmetric with respect to the $z=0$ plane, the other one asymmetric.  For the first, symmetric case 
we set $k_1=1.57$ and $k_2=0$. With these values the Mach number profile, which is shown in 
Figure\,\ref{fig:cfrag1}, has a 
very smooth, only mildly varying shape. Application of this Mach number profile to the MHS state results in the formation of quasi-3D 
tube-like current filaments. A selection of current isocontours of these filaments is depicted in the bottom panel of Figure\,\ref{fig:cfrag1}.
The current of this MHD flow has a 3D character, as can be seen in the middle panel of Figure\,\ref{fig:cfrag1} where we plot the current density
vector. The current density is strongest in the vicinity of the dipole field and around $z=0$.

For the second example, we use the following values for the constants in Equation\,(\ref{Mach_param}): $k_1=1.57$ and $k_2=0.75$. The resulting 
Mach number profile is depicted in the top panel of Figure\,\ref{fig:cfrag2}, the isocontours of the current filaments and the current density vector are shown in 
the bottom and middle panels of Figure\,\ref{fig:cfrag2}, respectively. The choice of $k_2\neq 0$ results in an asymmetry with respect
to the $z=0$ plane in the Mach number profile. In addition, the profile displays clearly stronger gradients. This leads to currents with 
very narrow, highly filamentary structures as can be seen in the image of the isocontours.

\subsection{2.5D scenario}

\begin{figure}[h]
\includegraphics[width=0.5\textwidth]{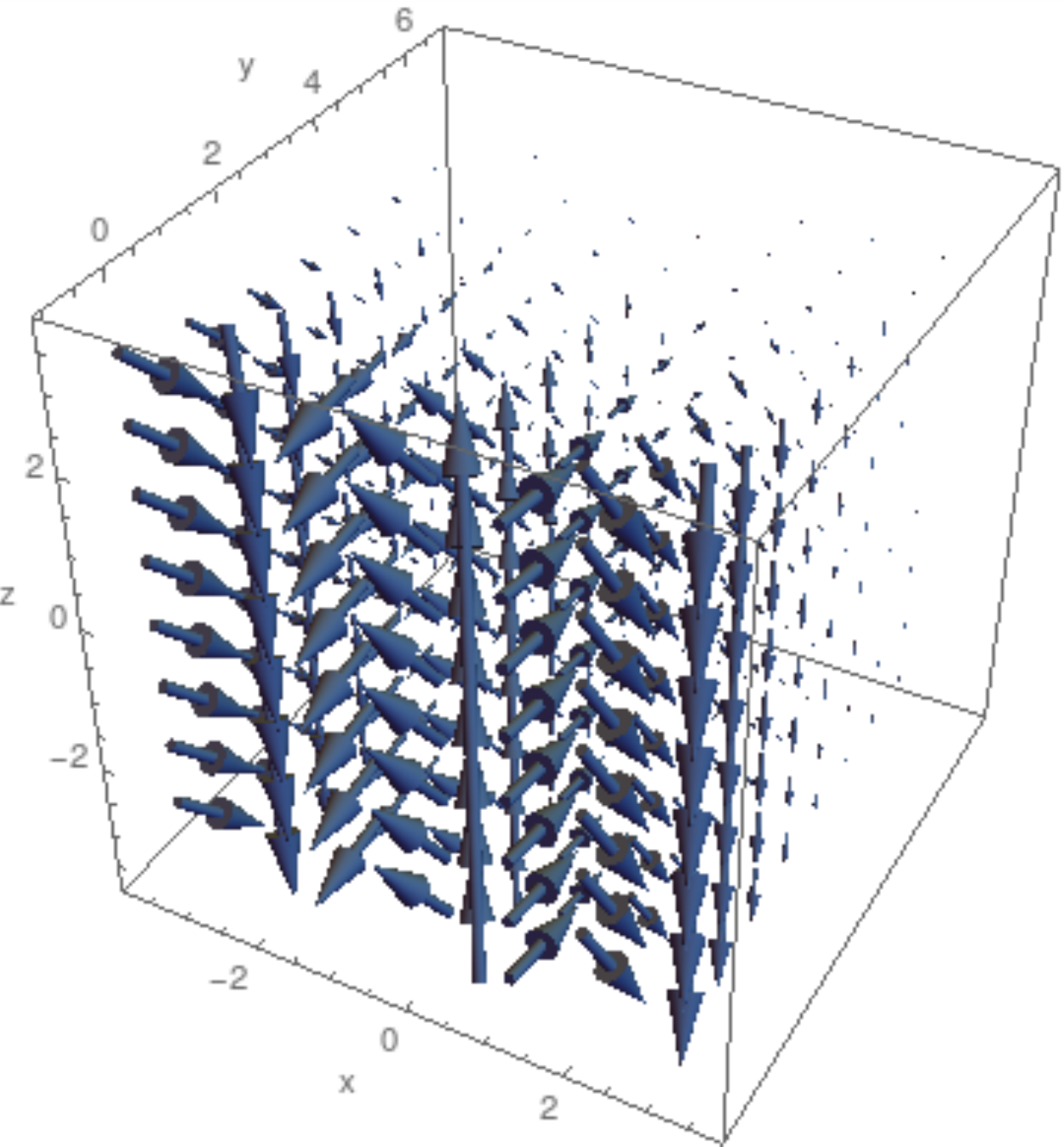}
\includegraphics[width=0.5\textwidth]{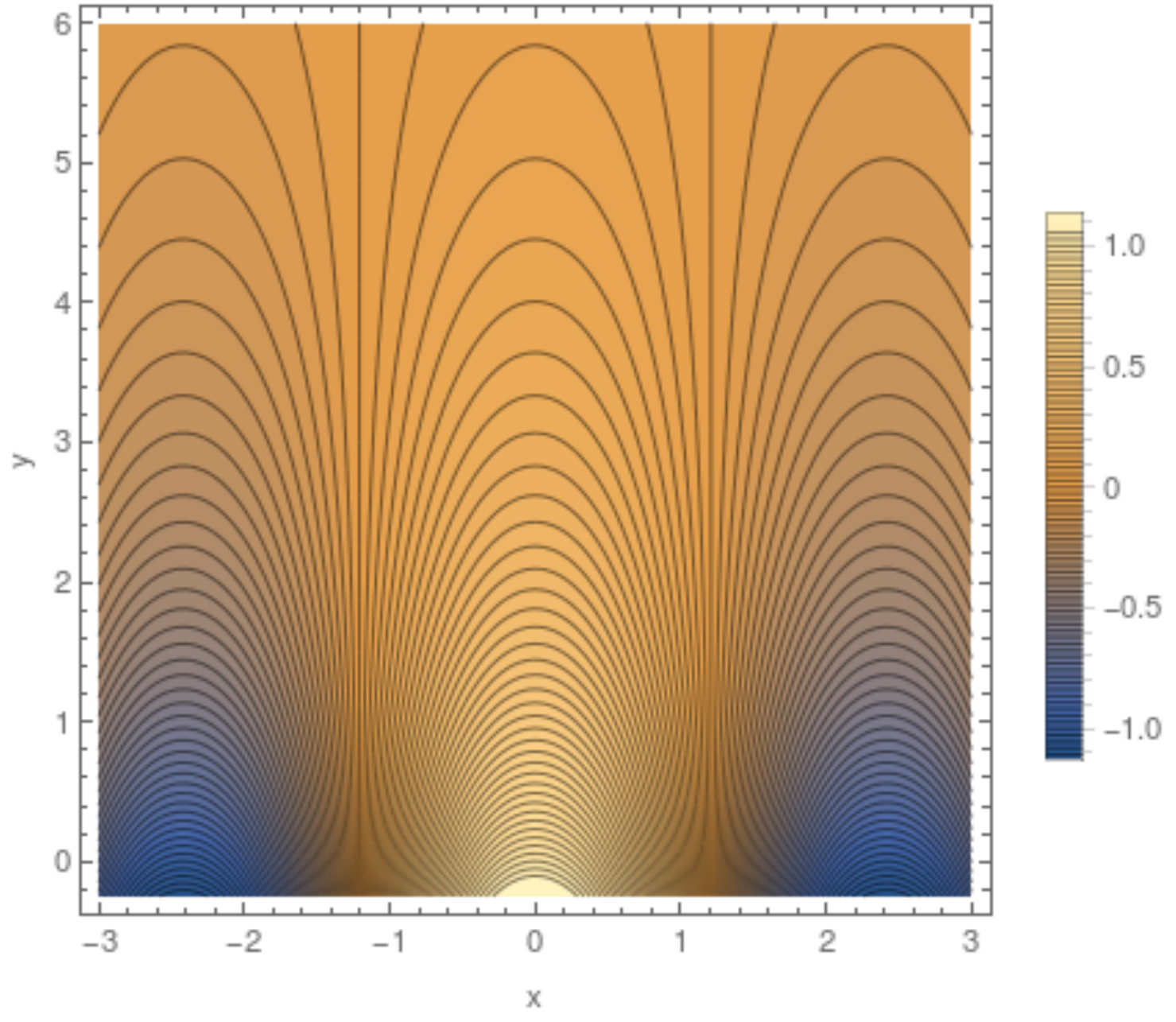}
\caption{Direction and strength of the initial force-free magnetic field (top) and the projection of the 
field lines of this field (bottom) into the $x$-$y$-plane in the 2.5D case. Color coding refers to
magnetic flux (function).}
\label{fig:cfrag3}
\end{figure}

For the 2.5D scenario we start from a linear force-free field as initial MHS equilibrium. Such fields are
typically chosen to model coronal magnetic fields \citep[see the review by][]{2012LRSP....9....5W}. 
We restrict to constant force-free fields, which means 
that the electric current density is given by $\textbf{\textit{j}} = c \textbf{\textit{B}}$ where $c$ is a constant,
and represent our force-free field with the 
Euler potentials $f$ and $g$ 
\begin{eqnarray}
\textbf{\textit{B}}_{S} &=& {\boldsymbol\nabla} f{\boldsymbol\times}{\boldsymbol\nabla} g \\
f &=& B_{0} \cos\left(k x\right) \exp\left(-\nu y\right)\\
g &=& z+\frac{c}{\nu} x
\, ,
\end{eqnarray}
where $\nu=\sqrt{k^2-c^2}$. For the presented case, we fix the constants at the following values:
$B_{0} = 1$, $k = 1.3$, and $c = 1.2$.

\begin{figure}[h]
\includegraphics[width=0.48\textwidth]{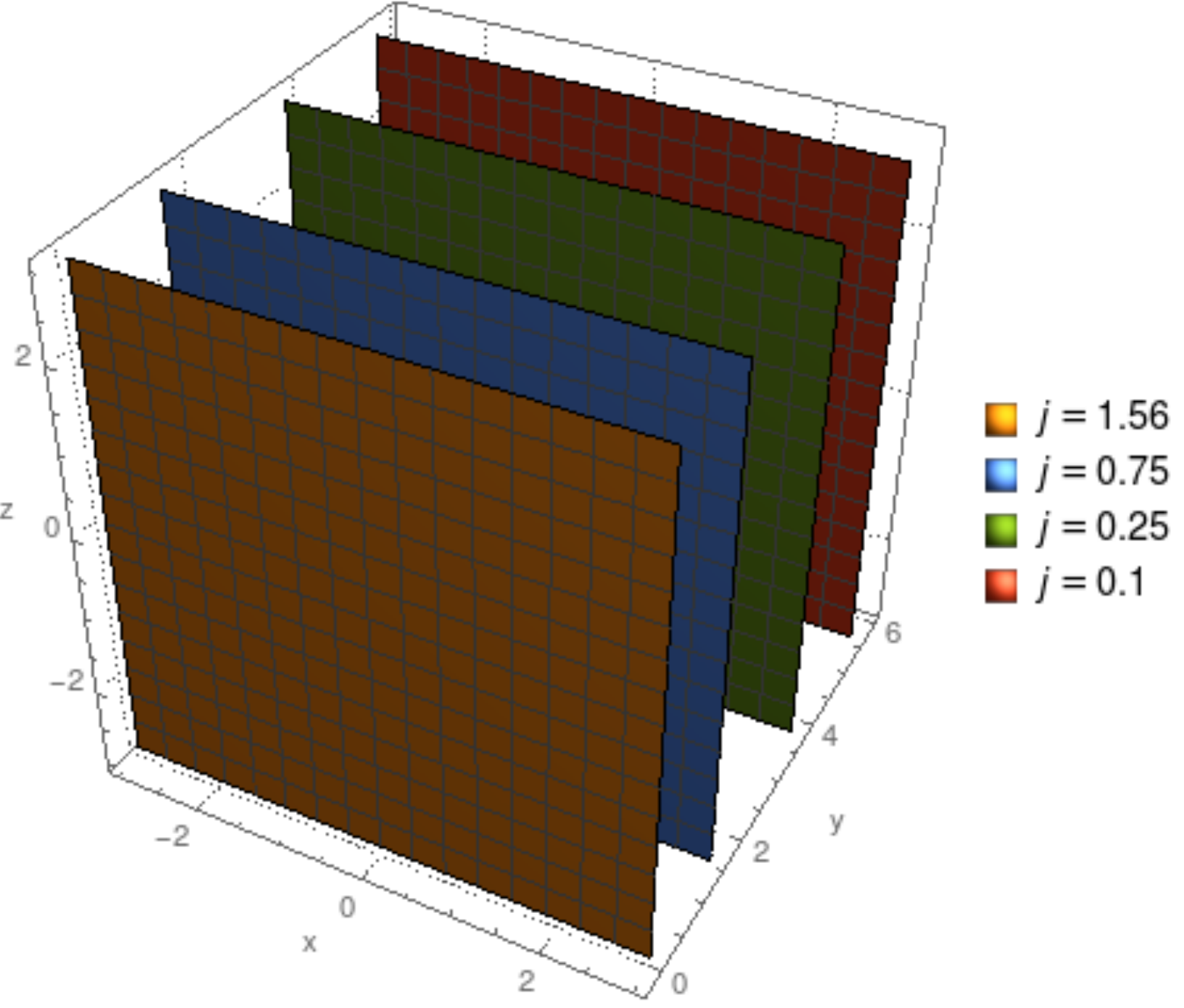}
\includegraphics[width=0.42\textwidth]{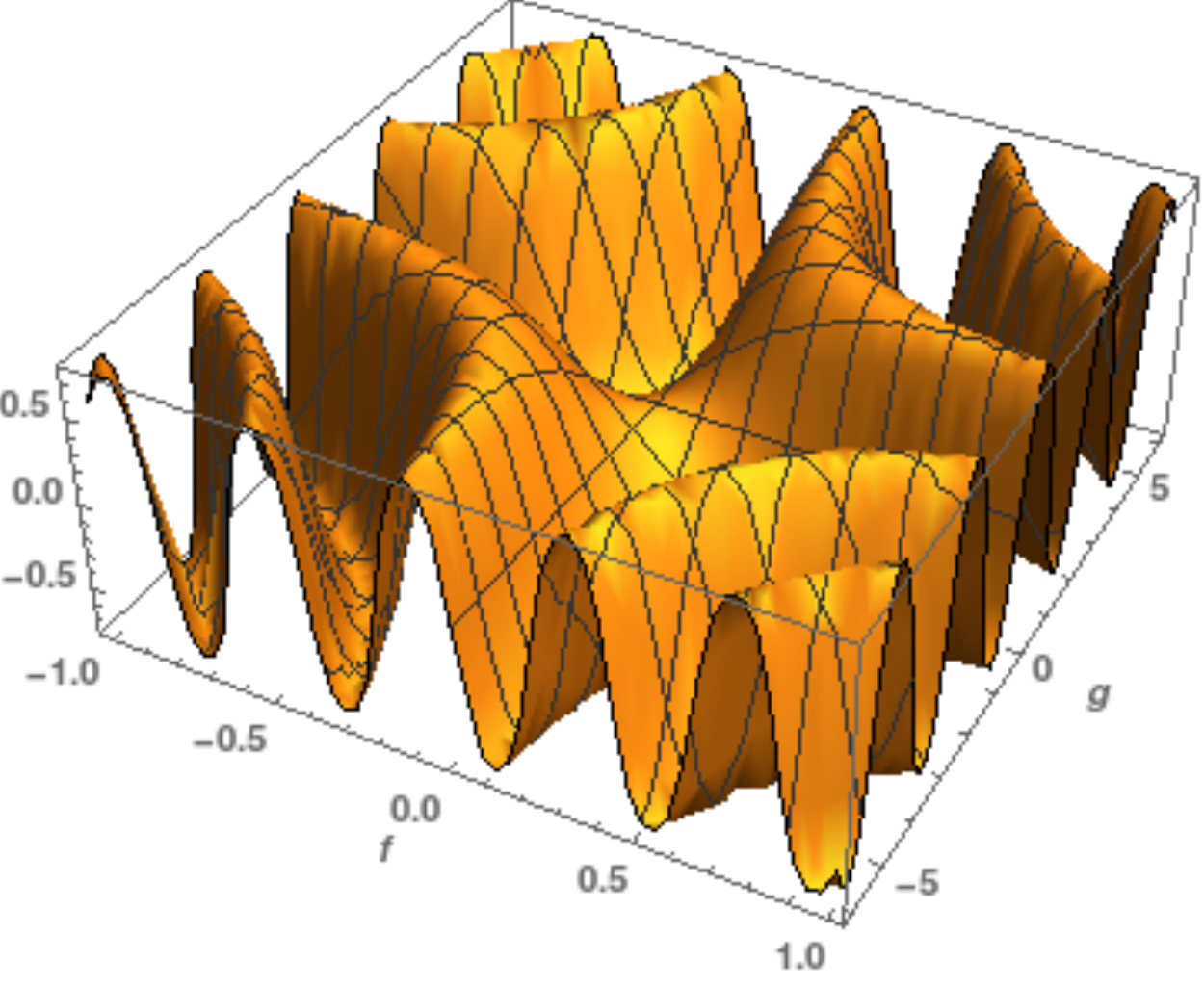}
\includegraphics[width=0.48\textwidth]{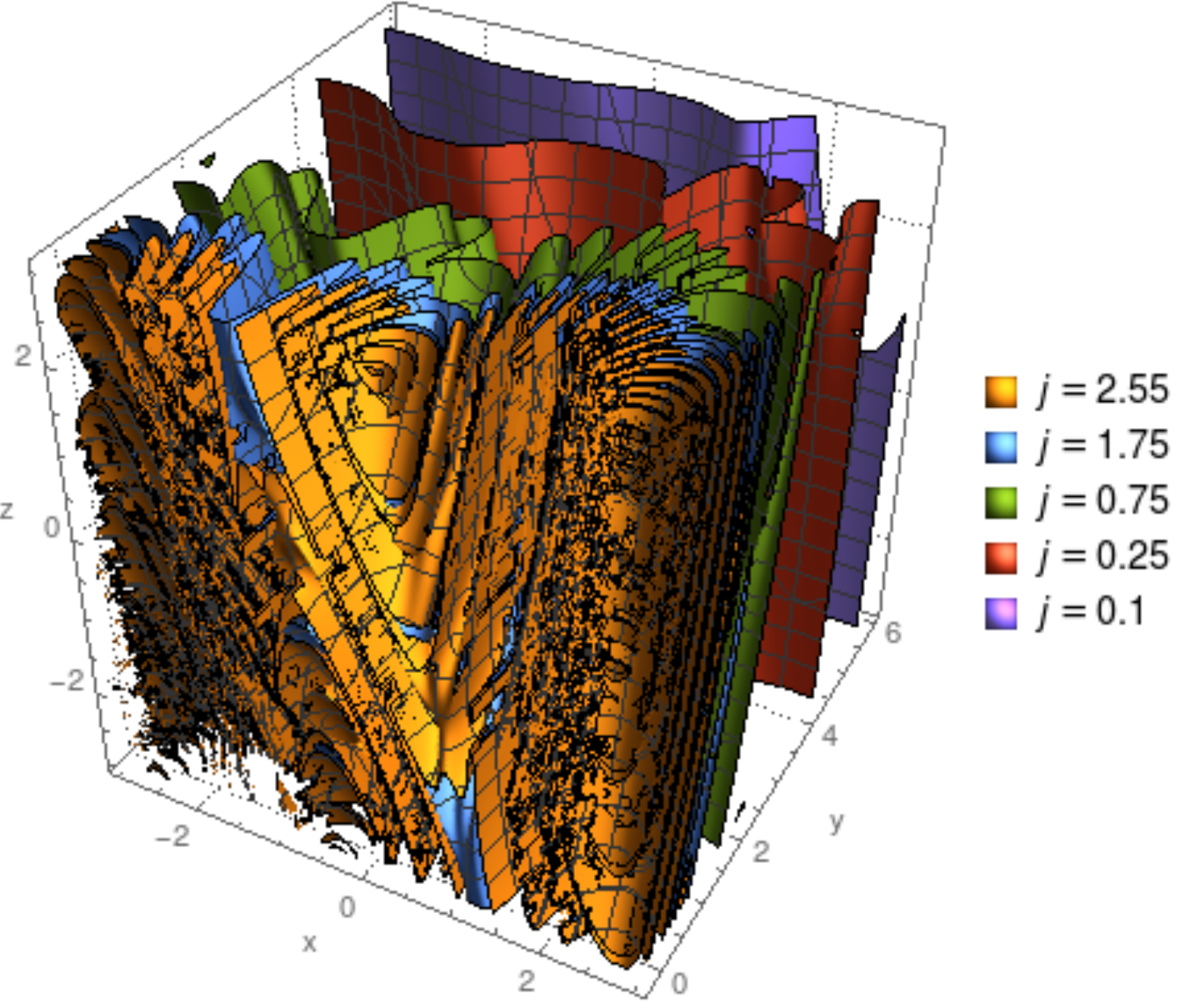}
\caption{Isosurfaces of the initial current density (top), Mach number profile (mid), and isosurfaces of
the transformed current density (bottom) in the 2.5D case.}
\label{fig:cfrag4}
\end{figure}

We chose the Euler potentials such that $f$ represents a component of the Fourier expansion of this 
force-free field \citep[see, e.g.,][]{1998PhST...74...77W}, and the second term of $g$ describes the 
component of the field in $z$-direction, i.e., the toroidal component, where $z$ is chosen as the 
invariant direction. The force-free magnetic field 
is shown in Figure\,\ref{fig:cfrag3}, where we plot its direction and strength (top panel) and the projection 
of the field lines into the $x$-$y$-plane (bottom panel). The magnetic field is strongest
for $y=0$ and decays with increasing values of $y$. Consequently, also the current density has its 
maximum at $y=0$. Moreover, with the chosen representation of the Euler potentials, the current
density is a pure function of $y$ and decays exponentially. Selected isocontours of this initial current
density are shown in the top panel of Figure\,\ref{fig:cfrag4}. Obviously, the isocontours are parallel to the 
$x$-$z$-plane and the maximum value of the current density is reached for $y=0$ where it has a numerical
value of 1.56.

We define the Mach number profile in the following form
\begin{equation}
M_{A} = \tanh \left[ \sin (2.5 f g ) \right]\, ,
\end{equation}
to provide a spatially strongly oscillating function. It is shown in the middle panel of 
Figure\,\ref{fig:cfrag4}. The results from applying this profile to the static equilibrium are shown
in the bottom panel of Figure\,\ref{fig:cfrag4}. Obviously, the former isocontour planes of the current 
density display now wavy structures with dependency in the initially invariant direction and their surfaces
are enlarged. Moreover, the numerical values of the mapped current density are larger, especially in the 
regions of high initial values of the magnetic field strength. These properties of the mapped current 
density (enlargement of both the isocontour surfaces and their numerical values) favor such kind of 
configurations for Ohmic dissipation.

\section{Discussion and Conclusions}

We present a general parametrization for the calculation of non-canonical transformations in the sub-Alfv\'{e}nic case. This parametrization
provides an ideal tool to calculate all possible transformations for a given MHS equilibrium, represented by the Euler potentials $f$ and $g$.
We apply this parametrization on 2D and 2.5D MHS equilibria and obtain symmetry breaking of the current, resulting in three current
components depending on all three spatial coordinates. The symmetry breaking implies that the magnetic field lines can have high symmetry and 
are ordered and non-chaotic (non-ergodic), but due to strong gradients of the flow the current distribution appears strongly shredded, displaying 
complex lamination. The additional fragmentation of the current filaments from Figure\,\ref{fig:cfrag1} into 
the highly filamentary structures 
seen in Figure\,\ref{fig:cfrag2} which is caused by only a slight change in the Mach number profile, shows 
that it is possible to obtain highly 
complex current distributions from an initially stationary and ordered magnetic field. These results 
demonstrate that to achieve strong currents, it is sufficient to have ordered fields and ordered flows. These 
currents are in principle suitable to trigger magnetic reconnection or pure
Ohmic heating. Moreover, our results imply that in contrast to Parker's idea of coronal heating, pure singular 
current sheets, i.e. tangential discontinuities, are not necessarily required. It is sufficient to have 
current sheets that are strong enough to overcome instability thresholds
for magnetic reconnection or to achieve required Ohmic heating rates. Such currents can easily be achieved 
with our model of symmetry breaking. However, it would be desirable to obtain those current density 
distributions which have sufficient strength and a suitable structure at the right locations to trigger
current driven instabilities. For this, an optimization procedure for the Mach number profile needs to be developed.

Another aspect of our studies is devoted to the question whether force-free fields are generic.
Although flows supporting force-free states have been a subject of investigation already in the
seventies \citep{1973Phy....67..323S, 1974Phy....78..321S}, only a limited set of such flows could be calculated.
However, these flows must obey very specific conditions, which means that the force-free state is not 
arbitrarily free, and even the force-free parameter $\alpha$ has to obey severe restrictions. As an example, 
\citet{1974Phy....78..321S} found that for axis-symmetric cases $\alpha$ must be a function of space and time,
while recent studies of \citet{2011PPCF...53i5013P} revealed that confined
solutions which necessarily need a monotonically decreasing pressure, cannot
exist.

Our analysis confirms and reinforces these previous findings. Moreover, it shows that force-free magnetic fields can be maintained by 
flows either only for specific geometries or for constant Mach numbers.

\acknowledgments
We thank the anonymous referee for his/her comments and suggestions.
This research has made use of NASA's Astrophysics Data System Bibliographic Services (ADS) and was 
supported from GA\,\v{C}R under grant numbers
16-05011S  
and
16-13277S. 
The Astronomical Institute Ond\v{r}ejov is supported by the project RVO:67985815.
This work was also partly supported by the European Union European Regional Development Fund,
project ``Benefits for Estonian Society from Space Research and Application'' (KOMEET, 
2014\,-\,2020.\,4.\,01.\,16\,-\,0029).

\bibliographystyle{aasjournal}
\bibliography{ms}



\clearpage

\end{document}